\documentclass[twocolumn]{article}
\usepackage[a4paper]{geometry}
\usepackage{amssymb}
\usepackage{euscript}
\usepackage{graphicx}
\graphicspath{{images/}}
\usepackage{color}
\usepackage{epsfig}
\newcommand{\image}[3]{
\begin{figure}[#1]
\begin{center}
\includegraphics{img_#2.pdf}
\caption{\small#3}
\label{image:#2}
\end{center}
\end{figure}
}

\def\ct{\cos\vartheta}
\def\st{\sin\vartheta}

\def\cpa{c_{\parallel}}
\def\cpe{c_{\perp}}

\usepackage[colorlinks=true,hypertexnames=true]{hyperref}
\usepackage[style=nature,backend=bibtex,doi=false,isbn=false,url=true,alldates=year]{biblatex}

\begin{document}
\title{Dynamics of $\vartheta$-solitons in the HPD state of superfluid $^3$He-B}

\author{V.~V.~Zavjalov\/\thanks{e-mail: v.zavjalov@lancaster.ac.uk}}

\date{23-12-2021}
\maketitle

\begin{abstract}
In this note we present numerical study of spin dynamics of
$\vartheta$-soliton in Homogeneously precessing domain (HPD) in
superfluid $^3$He-B. The soliton consists of a small heavy core and long
tails, in 1D simulation we observe their mutual motion. We also discuss
mass density and tension of the soliton membrane and its 3D oscillations
in a cylindrical cell.
\end{abstract}

\subsection*{Introduction}

Spin degrees of freedom in superfluid $^3$He-B have spontaneously broken
SO$_3$ symmetry~\cite{WV}. State of the system can be described by a
rotation matrix~$R({\bf n},\vartheta)$ with axis~$\bf n$ and rotation
angle~$\vartheta$. Spin-orbit interaction reduces the symmetry by fixing
angle~$\vartheta$ at value $\vartheta_L = \arccos(-1/4) \approx
\pm104^\circ$. In this system so-called $\vartheta$-solitons are possible
where $\vartheta$ changes between $\vartheta_L$ and $2\pi - \vartheta_L$.

In a usual NMR experiment $^3$He sample is placed in constant magnetic
field $H_z$ along $\bf\hat z$ axis and perpendicular radio-frequency (RF)
field $H_x$ rotating around~$H_z$ with angular velocity~$\omega$.
Depending on frequency shift~$\omega - \gamma H_z$ and amplitude of $H_x$
this system can be in two states: Homogeneously-precessing domain (HPD)
and Non-precessing domain (NPD) \cite{1985_hpd_e,1985_hpd_f_e}.
$\vartheta$-solitons can exist in both states. In NPD they have only a
small core where angle~$\vartheta$ is changing. Size of the core is about
10..40~$\mu$m, it is determined by so-called dipolar length (ratio of
gradient energy and energy of spin-orbit interaction). Analytical
description of solitons in NPD is quite simple~\cite{1976_maki_solitons}.
In HPD things are much more complicated. The core is surrounded by long
tails where $\vartheta$ has equilibrium value, but direction of
vector~$\bf n$ is changing. Theoretical discussion of solitons in HPD can
be found in~\cite{1992_mis_hpd_topol}. In experiments
$\vartheta$-solitons have been observed together with another type of
topological defect, spin-mass vortex, in
rotating~$^3$He-B~\cite{1992_prl_kondo_smv}.

\image{ht}{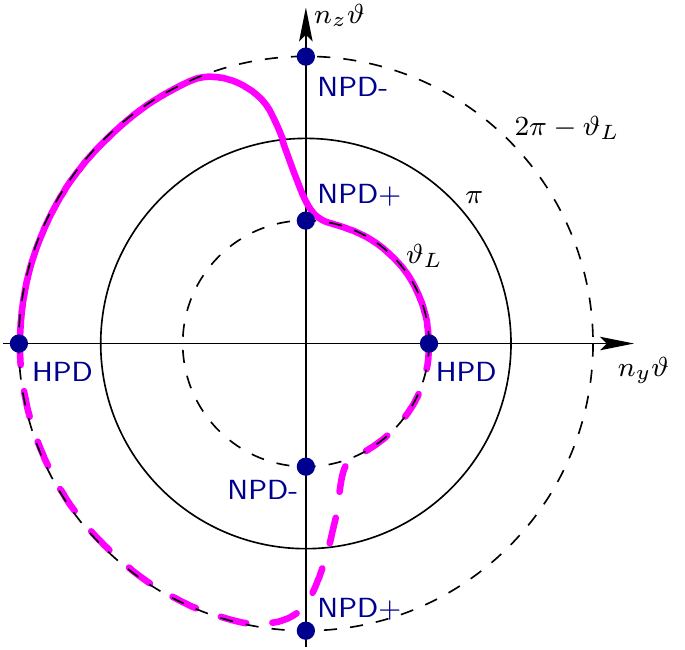}{Schematic representation of $\vartheta$-soliton in
$^3$He-B. See description in the text.}

On Fig.~1 a schematic picture of $\vartheta$-soliton in HPD is shown in
$\bf n\vartheta$ coordinates (vector with direction of $\bf n$ and
absolute value of $\vartheta$). Each state of rotation matrix can be
uniquely represented by a single point inside a sphere with radius $\pi$
(solid circle). Points outside the sphere can be mapped inside by adding
integer number of $2\pi$ rotations to $\vartheta$. Dashed circles
represent minimum of spin-orbit interaction, $\cos\vartheta = -1/4$.
Points NPD+, NPD-, and HPD are equilibrium states. Without RF field HPD
spontaneously breaks rotational symmetry and can exist in any point where
$n_z=0$, $\vartheta = \vartheta_L$, but presence of RF pumping along
$\hat{\bf x}$ stabilize HPD at $n_y = 1$. $\vartheta$-soliton in HPD is
schematically shown by thick magenta line. Dashed line is equivalent path
which can be obtained from the solid line by subtracting $2\pi$ from
value of~$\vartheta$. For simplicity picture is drawn in $y-z$ plane,
real solitons have a non-zero $n_x$ component.

This work is motivated by our experiments in 2017 in Helsinki (still
unpublished) where we observed unexpected oscillation modes in HPD. These
modes were localized in different parts of the experimental cell and had
frequencies increasing with frequency shift. One of suggestions was that
they are oscillations of $\vartheta$-solitons.

\subsection*{Numerical simulation of spin dynamics of $^3$He-B}

Dynamics of matrix~$R({\bf n},\vartheta)$ and spin $\bf S$ in superfluid
$^3$He-B is described by Leggett equations~\cite{1975_he3_teor_leggett}.
We write them in a frame rotating around~$\bf\hat z$ axis with angular
velocity~$\omega$ and solve 1D problem with only $z$-coordinate
dependence:
\begin{eqnarray}\label{eq:legg_rot}
\dot {\bf S} &=&
  {\bf S}\times (\gamma {\bf H} - \omega {\bf\hat z})
\\\nonumber
&& + {\bf T}^{D} + {\bf T}^{\nabla} + \mathcal{R}^{SD},
\\\nonumber
{\bf\dot n} &=&
-\frac12\ {\bf n}\times \Big [ 2\,\omega{\bf\hat z} + \Big(\frac{\gamma^2}{\chi} {\bf S} - \gamma {\bf H} \Big)
\\\nonumber
&&+ \frac{\st}{1-\ct}\ {\bf n}\times \Big(\frac{\gamma^2}{\chi} {\bf S} - \gamma {\bf H} \Big)
\Big],
\\\nonumber
\dot\vartheta &=& {\bf n} \cdot \Big(\frac{\gamma^2}{\chi} {\bf S} - \gamma {\bf H} \Big) + \mathcal{R}^{LT}.
\end{eqnarray}
Here $\gamma$ is gyromagnetic ratio of~$^3$He, $\bf H$ is external
magnetic field, $\chi$ is magnetic susceptibility.
We use magnetic field which is constant in the rotating frame:
\begin{equation}
\gamma {\bf H} = \gamma H_x {\bf\hat x} + \gamma H_z {\bf\hat z} = \omega_x {\bf\hat x} + \omega_z {\bf\hat z},
\end{equation}
and introduce frequency shift
$\delta\omega = 2\pi\,\delta f = \omega_z - \omega$.

Dipolar torque~${\bf T}^D$ comes from spin-orbit interaction and can be written as
\begin{eqnarray}\label{eq:td_nt}
{\bf T}^{D} = \frac{4 \Omega_B^2\chi}{15\gamma^2}\ \st(4\ct + 1)\ {\bf n},
\end{eqnarray}
where strength of the interaction is represented by parameter $\Omega_B$,
Leggett frequency.

Gradient torque~${\bf T}^{\nabla}$ can be written as
\begin{eqnarray}
\nonumber
T_a^{\nabla} &=&
\frac{\chi}{\gamma^2}\ e_{abc} R_{cj}
\Big[ (\cpe^2 - \cpa^2/2) R_{bj}''
\\
&+&
2(\cpa^2-\cpe^2) R_{b3}''\delta_{j3}\Big],
\end{eqnarray}
where $\cpa$ and $\cpe$ are spin-wave
velocities~\cite{2015_jetp_zavj_spinwaves}, and primes are derivatives
over $z$ coordinate. Expression for gradient torque via~$\bf n$ and
$\vartheta$ is huge, we are not writing it here.

There are two relaxation terms: spin diffusion~$\mathcal{R}^{SD}$
and Leggett-Takagi relaxation~$\mathcal{R}^{LT}$:
\begin{eqnarray}
\mathcal{R}^{SD} &=& D\ {\bf S}'',
\\\nonumber
\mathcal{R}^{LT} &=& \frac4{15}\ \st(4\ct + 1)\ \frac1{\tau}
\end{eqnarray}
with two additional parameters: spin diffusion coefficient~$D$ and
Leggett-Takagi relaxation time~$\tau$. In general, spin diffusion
coefficient is a tensor, but for 1D problem only a single component
$D=D_{zz}$ enters the formula (see~\cite{1985_hpd_f_e}).

We use PDECOL algorithm~\cite{1979_PDECOL,1992_PDECOL} for integrating
equations~(\ref{eq:legg_rot}) numerically. Original program have been
written a long time ago by V.Dmitriev, we made a few
modifications~\cite{vmcw} for this work: Instead of using seven
coordinates ($\bf S$, $\bf n$, and $\vartheta$) we use six, $\bf S$ and
$\vartheta {\bf n}$. By doing this we avoid additional constraint $|{\bf
n}|=1$, uncertainty at $\sin\vartheta=0$, and unphysical difference
between ($\vartheta$, $\bf n$) and ($-\vartheta$, $-\bf n$) states. We
use non-uniform adaptive mesh in the calculation, which is important
because solitons have two different scales: micron-size core and
millimeter-size tails. To stabilize the soliton we use exact solution for
soliton core in NPD as initial conditions. It can be obtained from Leggett
equations by setting all time derivatives to zero and using ${\bf n} =
{\bf\hat z}$. This leads to a differential equation:
\begin{equation}\label{eq:th_sol_npd0}
\frac{12}{13}\xi_D^2\ \vartheta'' = -\st(4\ct + 1),
\end{equation}
where a conventional definition of dipolar length $\xi_D^2 =
\frac{65}{16}(2\cpe^2-\cpa^2)/\Omega_B^2$ is used. Solving it with appropriate
boundary conditions we get soliton profile:
\begin{equation}\label{eq:th_sol_npd}
\vartheta = 2 \tan^{-1}\left[\sqrt{\frac{3}{5}} \tanh \left(\sqrt{\frac{65}{64}}\ \frac{\pm z}{\xi_D}\right)\right] + \pi.
\end{equation}
One more problem we met in calculation was mobility of the soliton. To
localize it we use a small recess in Leggett frequency profile, 5-10\%
decrease at $\xi_D$ width. Soliton with its large spin-orbit energy is
pinned by this feature. By varying pinning strength we can check how this
method affects calculation results. We see that there is no noticeable
effect on soliton oscillation frequency and only a small change (${<}1$\%) in
soliton mass calculation.

We use frequency 1.124~MHz, and cell length 0.9~cm, same as we had in
the experiment. The soliton is created in the middle of the cell and
stabilized by waiting long enough time, changing relaxation terms, or
doing averaging of motion over some time and using result as new initial
conditions. Starting from a stable state with the soliton we excite
oscillations by a small change of frequency shift (which is equivalent to
a small step in static magnetic field) and record components of integral
magnetization. Fourier spectrum of magnetization is fitted using
a model signal which is sum of a few decaying harmonic oscillations.
We observed both oscillations of the soliton and uniform
HPD oscillations~\cite{2005_he3b_hpd_osc}. Frequency of uniform
oscillations is known, it can be used as an additional check of
calculation accuracy:
\begin{equation}\label{eq:hpd_osc}
\Omega^2 = \frac{4}{\sqrt{15}}\ \frac{\omega_x\omega\,\Omega_B^2}{\omega^2+8/3\,\Omega_B^2}.
\end{equation}
With this excitation method we usually observe two first modes of
soliton oscillations. Third and forth modes are much weaker and can be
seen only at some conditions.

\subsection*{Results}
\image{ht}{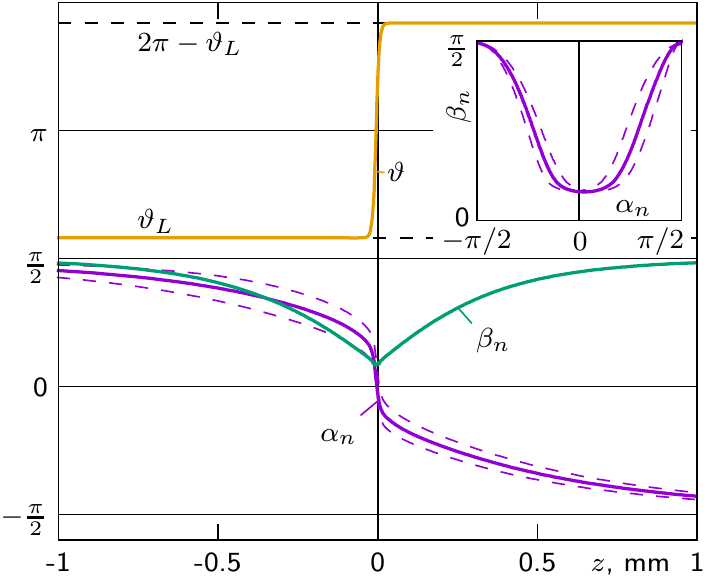}{Soliton profile: angle $\vartheta$,
polar ($\beta_n$) and azimuthal~($\alpha_n$) angles of~$\bf n$ vs. coordinate $z$.
Dashed color lines near $\alpha_n$ profile show main mode of soliton oscillations.
Inset: $\beta_n$ vs. $\alpha_n$ plot. The figure is
calculated with $c_{\perp}=1115$~cm/s, $\Omega_B/2\pi=235.8$~Hz (which
corresponds to $T=0.6~Tc, P=25.7$~bar), $D=0.01$~cm$^2$/s, no
Leggett-Takagi relaxation, $H_x = 4$~mG, frequency shift $\delta f$ =
40~Hz.}

On Fig.~2 an example of soliton equilibrium profile is shown. Direction
of vector~$\bf n$ is represented by polar and azimuthal angles,
$\beta_n$, and~$\alpha_n$. One can see a small core where
angle~$\vartheta$ changes between equilibrium values~$\vartheta_L$
and~$2\pi-\vartheta_L$ at distance~$\sim 10\mu$m and tails where $\bf n$
is rotated at distance~$\sim 1$~mm. It's interesting to see that soliton
tails are slightly asymmetric (they go towards different states, NPD+ and
NPD- ), and that in the center vector $\bf n$ does not reach vertical
position as it was suggested in~\cite{1992_mis_hpd_topol}. Soliton core
is not just change of~$\vartheta$, it also includes rotation of~$\bf n$
by some noticeable angle. Main mode of soliton oscillations is shown by
dashed color lines. In this mode the soliton tail is rotating around
vertical axis, only azimuthal angle~$\alpha_n$ is changing.

\image{ht}{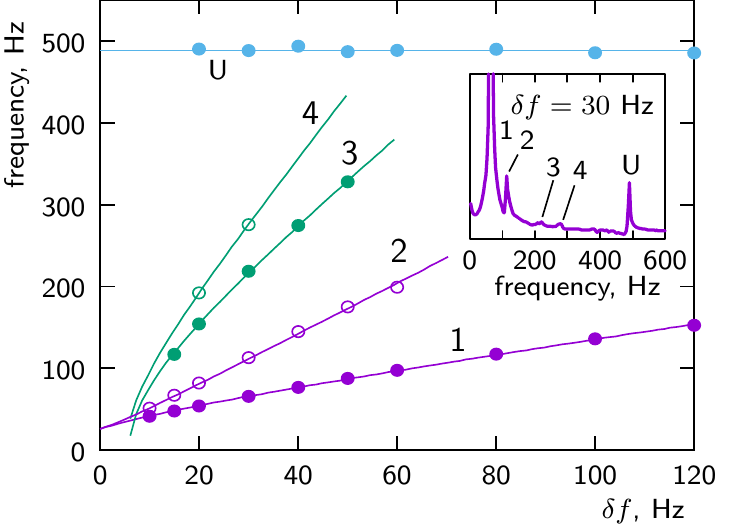}{Oscillation frequencies as a function of frequency
shift. Calculation is done for $\Omega_B/2\pi = 100$~kHz, $c_\perp =
1000$~cm/s, $H_x=8$~mG. One can see four soliton oscillation modes
(labeled ``1''..``4'') and HPD uniform mode (labeled ``U''). Inset shows
an example of magnetization spectrum at frequency shift $\delta f =
40$~Hz. Lines are fits to formula~(\ref{eq:fre}) with free parameters $B$
and $C$.}

\image{ht}{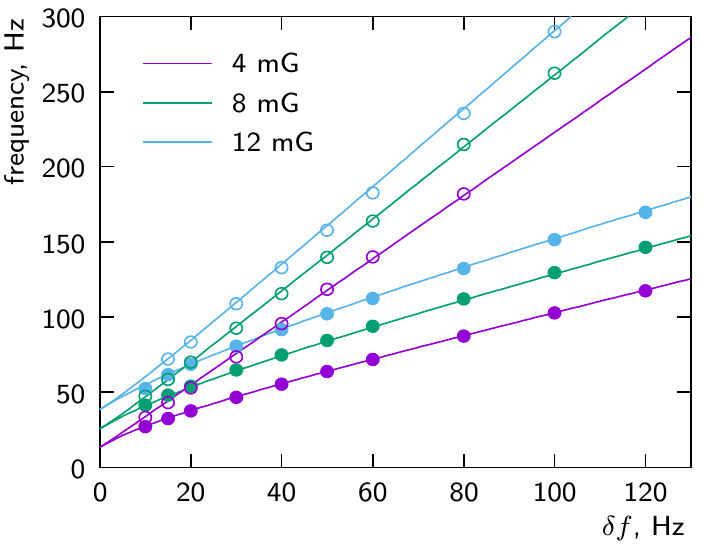}{Frequencies of soliton modes 1 and 2 as a function of
frequency shift for three different values of RF pumping: 4~mG (magenta
points), 8~mG (green points), 12~mG (blue points). Calculation is done
for $T=0.6\,T_c$, $P=25.7$~bar ($\Omega_B/2\pi = 235.8 $~kHz, $c_\perp =
1115$~cm/s). Lines are drawn using approximation formulas~(\ref{eq:fre})
and~(\ref{eq:fre_bc}).}

On Fig.~3 oscillation frequencies are shown as a a function of frequency
shift. The highest, shift-independent mode is uniform HPD
oscillations~(\ref{eq:hpd_osc}), others are oscillations of the soliton.
We found that frequencies of all soliton modes as a function of frequency
shift~$\delta\omega$ can be written as:
\begin{eqnarray}\label{eq:fre}
\Omega^2 &=& B\,\omega(\delta\omega\pm\omega_x/\sqrt{15}) + C\,\delta\omega^2
\end{eqnarray}
with ``$+$'' sign for modes 1, 2 and ``$-$'' for modes 3, 4. There is no
exact theory behind this formula, it's just an approximation which works
for all observed modes. Note that ratio $\sqrt{15}$ between RF pumping
and frequency shift is special for HPD: at $\delta\omega <
\omega_x/\sqrt{15}$ uniform HPD state becomes unstable. Parameters $B$
and $C$ depend only on RF pumping amplitude~$\omega_x$ and Leggett
frequency~$\Omega_B$. Me have measured modes 1 and 2 for $P=25.7$~bar,
$T=0.6\ldots0.9 T_c$, $H_x = 4\ldots12$~mG and found approximation
formulas (see Fig.~4):
\begin{eqnarray}\label{eq:fre_bc}
B_{1,2} &=&\sqrt{\frac{\omega_x}{\omega}}
  \left(\frac{1}{107} + 372\frac{\omega_x}{\omega}\right),
\\\nonumber
C_1 &=& 0.278 + 27.4\sqrt{\frac{\omega_x}{\Omega_B}}
  \left(1 + 0.04 \frac{\omega}{\Omega_B}\right),
\\\nonumber
C_2 &=& 1.16  + 319\sqrt{\frac{\omega_x}{\Omega_B}}
  \left(1 + 0.08 \frac{\omega}{\Omega_B}\right).
\end{eqnarray}
We also found that relaxation of soliton oscillations
is determined by spin diffusion, but not Leggett-Takagi relaxation.

\subsection*{Oscillation of a soliton membrane}

Let's calculate mass density, associated with a soliton. First we solve
second and third Leggett equations~(\ref{eq:legg_rot}) for spin:
\begin{eqnarray}\label{eq:s_nt}
\frac{\gamma^2}{\chi} {\bf S} &=& \gamma {\bf H} + {\bf n}\ \dot\vartheta
\\\nonumber
&+& (1-\ct)\ {\bf n}\times ({\bf\dot n} + {\bf n}\times\omega {\bf\hat z})
\\\nonumber
&+& \st\ ({\bf\dot n} + {\bf n}\times \omega {\bf\hat z}).
\end{eqnarray}
From this formula one can see how spin is created by magnetic field and
by motion of~${\bf n}$ and $\vartheta$. There are also terms with
$\omega$ which come from rotation of the frame.
Consider a soliton which is moving with velocity~$v$ along~${\bf\hat z}$
axis. We can write ${\bf\dot n} = \frac{d}{dt} {\bf n}(z + v\,dt) = v\,{\bf n}'$,
and same for $\vartheta$. Using formula~(\ref{eq:s_nt}) we can
find spin created by soliton motion. Energy associated with spin is
$F = \frac{\gamma^2}{2\chi}{\bf S}^2 - ({\bf S} \cdot \gamma {\bf H})$.
Term proportional to $v^2$ will give us mass density of the soliton:
\begin{equation}\label{eq:mass}
m = \frac{\chi}{\gamma^2} \left[
(\vartheta')^2 + 2(1-\ct)\ ({\bf n'})^2
\right]
\end{equation}
In this discussion we assumed that moving soliton have same profile as
stationary one. This is true for small velocities. Formula for a
soliton moving with arbitrary velocity can be found
in~\cite{1986_rozhkov_solitons}.

\image{ht}{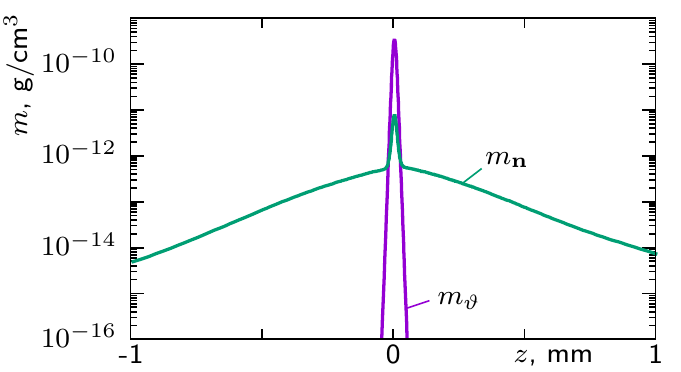}{Mass profile of a soliton. Total mass is split
into two parts, corresponding to two terms of~(\ref{eq:mass}). Calculated
with same parameters as Fig.~2.}

\image{ht}{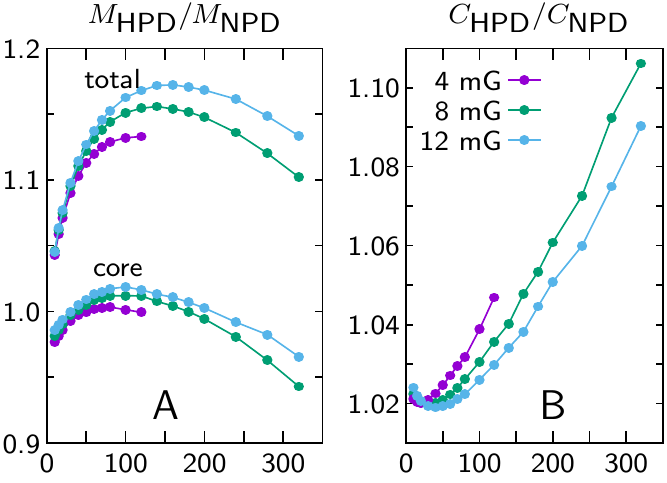}{{\bf A:} Ratio of soliton mass in HPD (total mass
and mass of the core) and in NPD as a function of
frequency shift, for three different values of RF pumping: 4~mG (magenta
points), 8~mG (green points), 12~mG (blue points). {\bf B:} Ratio of
wave velocity~$C$ for soliton in HPD and in NPD.}

Consider a $\vartheta$-soliton in
NPD~(\ref{eq:th_sol_npd0}-\ref{eq:th_sol_npd}) pinned to walls of a
cylindrical cell perpendicular to cell axis. It can move as a circular
membrane. We can find mass~$M$ of unit area by integrating
(\ref{eq:mass}) along $\bf\hat z$ axis. We calculate tension~$T$ of the
membrane as total potential energy (gradient and spin-orbit) of the soliton per
unit area. We ignore additional gradient forces which come from bending of the
soliton because they are small if cell size is bigger then soliton thickness. Using
soliton profile~(\ref{eq:th_sol_npd}) we obtain
\begin{equation}\label{eq:npd_mass}
M = \frac{13}{3}\frac{\chi}{\gamma^2\xi_D^2}
\int_{-\infty}^{+\infty}\Big(\ct(z) + \frac14\Big)^2 dz,
\end{equation}
\begin{equation}
T = \frac{16}{15} \frac{ \chi\Omega_B^2}{\gamma^2}
\int_{-\infty}^{+\infty}\Big(\ct(z) + \frac14\Big)^2 dz.
\end{equation}
Ratio of tension and mass give square of wave velocity~$C$ along the membrane,
\begin{equation}\label{eq:npd_c}
C^2 = 2\cpe^2-\cpa^2.
\end{equation}
Oscillation frequencies of the membrane are determined by zeros of Bessel
functions, the first one for membrane radius~$r$ is $2.405\,C/r$.
For $T=0.6~T_c$, $P=25.7$~bar, and $r=3.9$~mm this gives 1.7~kHz.

In numerical simulation we can find mass and tension of the soliton in
HPD and compare it with NPD case. It is also possible to separate masses
of tail and core of the soliton. On Fig.~5 example of mass density of a
soliton in HPD is presented. Mass is splitted into two parts which
correspond to two terms in~(\ref{eq:mass}): $m_{\vartheta}$ comes from
motion of $\vartheta$, and $m_{\bf n}$ is created by motion of~$\bf n$.
On Fig.~6 calculated mass and wave velocity of soliton in HPD is compared with
values for soliton in NPD (formulas \ref{eq:npd_mass} and \ref{eq:npd_c}).

% TODO - correction for pinning potential

\subsection*{Conclusions}

Homogeneously precessing domain is a well-defined system which can be
used as a sensitive tool for studying $^3$He properties. Understanding of
topological defects in this system should be important for future
experiments. In this work we developed numerical simulation of
$\theta$-solitons in HPD. We observed a few oscillation modes of soliton
tails with respect to its core. We also discuss 3D oscillations of the
soliton membrane pinned to walls in both HPD and NPD. One can imagine
other oscillation modes in this system, for example non-uniform motion of
tails in the plane of the soliton.

We think that further theoretical, numerical, and experimental
investigation of $\theta$-solitons in HPD is needed. It is still not
clear what is reliable way of producing solitons in experiment, and were
oscillation modes observed in experiments mentioned above
related to solitons.

\onecolumn

\printbibliography

\end{document}